\documentclass[10pt,conference,compsocconf]{IEEEtran}
\usepackage{times}
\usepackage[sort,nocompress]{cite}

\usepackage{caption}
\usepackage[ruled,linesnumbered,lined,commentsnumbered]{algorithm2e}
\usepackage[tight]{subfigure}
\usepackage{graphicx}
\usepackage{algorithmic}
\usepackage{booktabs}

\ifCLASSINFOpdf
\else
\fi
\begin{document}
\pagenumbering{gobble}
%
\title{\textbf{\Large Taming Near Repeat Calculation for Crime Analysis via Cohesive Subgraph Computing}\\[0.2ex]}

\author{\IEEEauthorblockN{~\\[-0.4ex]\large Zhaoming Yin\\[0.3ex]\normalsize}
\IEEEauthorblockA{Open Data Processing Platform Team,\\
Alibaba Cloud\\
Hangzhou, Zhejiang, China\\
Email: {\tt stplaydog@gmail.com}}
\and
\IEEEauthorblockN{~\\[-0.4ex]\large Xuan Shi\\[0.3ex]\normalsize}
\IEEEauthorblockA{Department of Geosciences\\
University of Arkansas\\
Fayetteville, Washington County, USA \\
Email: {\tt xuanshi@uark.edu}}
}


%


\makeatletter
\newcommand{\removelatexerror}{\let\@latex@error\@gobble}
\makeatother

\maketitle

\begin{abstract}
Near Repeat (NR) is a well-known phenomenon in crime analysis, assuming that crime events exhibit correlations within a given time and space frame. Traditional NR calculation would generate two event pairs if two events happened within a given space and time limit. When the number of events is significant, however, NR calculation is time consuming and how these pairs are organized has not yet been explored. In this paper, we designed a new approach to calculate clusters of NR events efficiently. To begin with, R-tree is utilized to index crime events. A single event is represented by a vertex, whereas edges are constructed by range-querying the vertex in R-tree; this way, a graph is formed. Cohesive subgraph approaches are applied to identify the event chains. k-clique, k-truss, k-core plus Density-based Spatial Clustering of Applications with Noise (DBSCAN) algorithms are implemented in sequence to their varied range of abilities to find cohesive subgraphs. Real-world crime data in Chicago, New York, and Washington DC are utilized to conduct experiments. The experiments confirmed that near repeat has a substantial effect on real big crime data by conducting Map-reduce empowered Knox tests. The performances of 4 different algorithms are validated, with the quality gauged by the distribution of the number of cohesive subgraphs and their clustering coefficients. The proposed framework is the first to process the real crime data of million records and is the first to detect NR events with a size of more than 2.

\end{abstract}



%
\IEEEpeerreviewmaketitle

\section{Introduction}
In criminal research, it was found that when a crime incident takes place at a given geographical location, its neighboring areas would have a higher possibility of experiencing follow-up incidents in a short period \cite{ratcliffe2008near}\cite{short2009measuring}\cite{wells2011patterns}. When the first incident occurs at a specific time, the follow-up incident at the same location and close to the initial time is a repeat. The incidents that occur near the space and time of the initiator are called near-repeat. Such repeat and near-repeat phenomena have been found from burglaries and gun violence studies, and have important implications to dispatch police force in crime mitigation activities \cite{ratcliffe2008near}\cite{townsley2007near}\cite{wells2011patterns}. To prove the near-repeat effect, the classic way is to use the Knox test method \cite{ratcliffe2008near}\cite{short2009measuring}\cite{townsley2003infectious}. The general idea of Knox test is to calculate the pairwise distance (in terms of space and time) between different crime events, and place the event pair into different bins of a table; the residual value of each specific entry of the table is calculated to indicate how random these pairs are organized into the range. The issue with this method is that its time complexity is $O(n^2)$ ($n$ is the number of crime events). When dealing with big real-world data, it will take days, if not months, to finish the computing task.

When near-repeat research only considers the space-time interaction among every two incidents, a complete “space-time event chain” is more appropriate to differentiate such a scenario of separate space-time pairs \cite{townsley2007near}. For example, when three shooting events (A and B), (B and C), (A and C) comply with the near-repeat definition, a three-event chain can be identified. In this case, it would be more meaningful to identify the correlation between multiple incidents rather than just two. Event chain analysis improves our understanding of the role of space and time among series of shooting events, or other types of crime events. The significant existence of paired shooting events does not mean the meaningful presence of multiple shooting event chains in the same space-time context. Otherwise, all initiators or follow-up shooting events should be close to each other and form a spatial cluster in a city.

Enumerating event chains in a brute-force way would be extremely difficult because the time complexity grows exponentially. Nevertheless, we can abstract the problems of near-repeat event chain detection by dividing it into two separate issues: 1) detecting near repeat pairs efficiently; 2) clustering or chaining near repeat pairs with high speed.

To begin with, the most efficient way to avoid unnecessary pairwise computation of each crime event is to use an index to organize events, such that one only needs to query its spatial-temporal adjacent events to generate event pairs, and R-tree \cite{guttman1984r} is the right choice. Once all event pairs are detected, they can be represented as a graph.

Furthermore, detecting a chain of near-repeat events can be modeled as a cohesive subgraph enumeration problem \cite{cohen2008trusses}. Ideally, all events should have connections between each other within a cohesive subgraph, and such a subgraph is a k-clique ($k$ is the number of vertices in the subgraph) \cite{konc2007improved}; Nevertheless, in the real world, the graph is massive, and approximation methods are more efficient than exact algorithms \cite{satish2014navigating}. Rather than asking all vertices in a subgraph to have a connection between each other, a k-core \cite{batagelj2003m} only asks that each vertex in the subgraph has k degrees, and this restriction is relaxed. A variant of Density-based Spatial Clustering of Applications with Noise (DBSCAN) algorithm \cite{birant2007st} can be applied to detect clusters of spatial-temporal data. The DBSCAN algorithm is fast with a complexity of $O(Vlog(V))$. One of the alternative versions requires that each edge in the subgraph should be in $k-2$ triangles; this is called k-truss \cite{cohen2008trusses}. In recent years, lots of advances had been made in the area of truss decomposition, regarding speed \cite{wang2012truss}, a variance of graph \cite{huang2016truss}, and data streaming \cite{huang2014querying}, which make the k-truss algorithm a preferred alternative to the k-clique method. All three alternatives to the k-clique algorithm have polynomial complexity.

{\bf Organization}.
In this paper, we will discuss a framework to incorporate the methods of indexing crime events, computing event pairs, and detecting near-repeat event chains applying k-clique,  k-core, DBSCAN, and k-truss algorithms separately. Section 2 formally defines the near repeat chain detection problem, gives the necessary notations, and describes the framework of our near repeat chain detection methods; Section 3 reports the experimental results using real-world data; the Conclusion is derived in the last section of this paper.

\section{Event Chain Calculation through Graph Analytics}
\subsection{Using graph to represent Crime Events}
\begin{figure*}[htbp!]
    \begin{center}
	    \subfigure[Origin]{
            \label{origin}
            \includegraphics[width=0.25\textwidth]{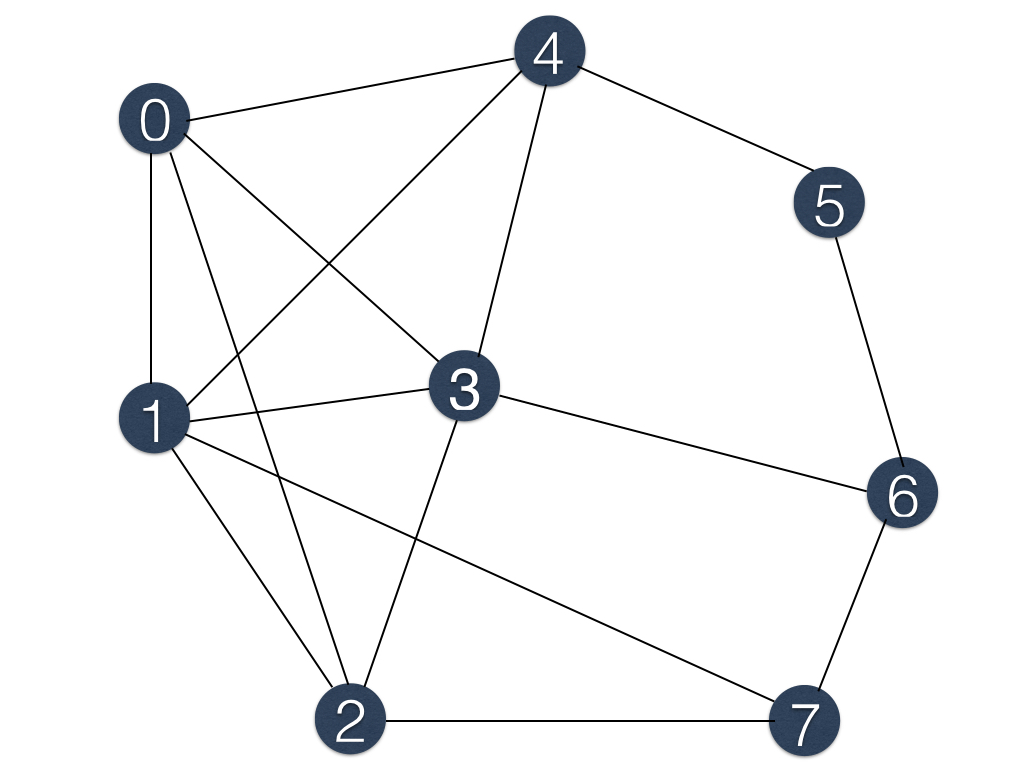}
        }%
	    \subfigure[4-clique]{
            \label{3_clique}
            \includegraphics[width=0.15\textwidth]{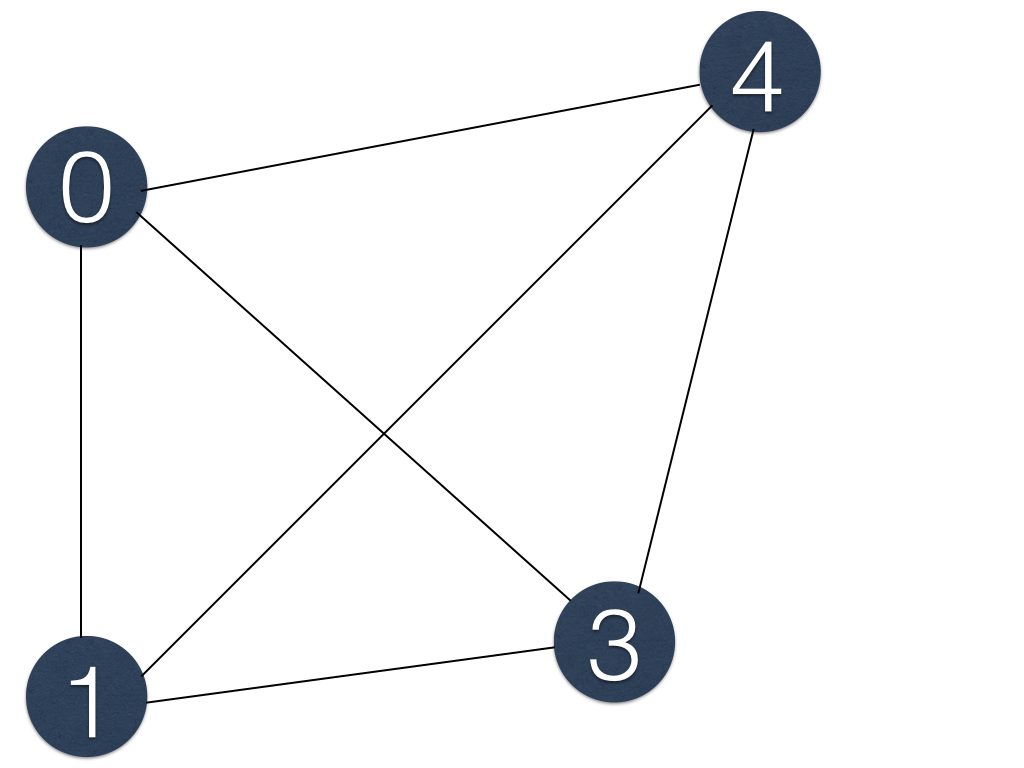}
        }%
        \subfigure[3-core]{
           \label{3_core}
           \includegraphics[width=0.25\textwidth]{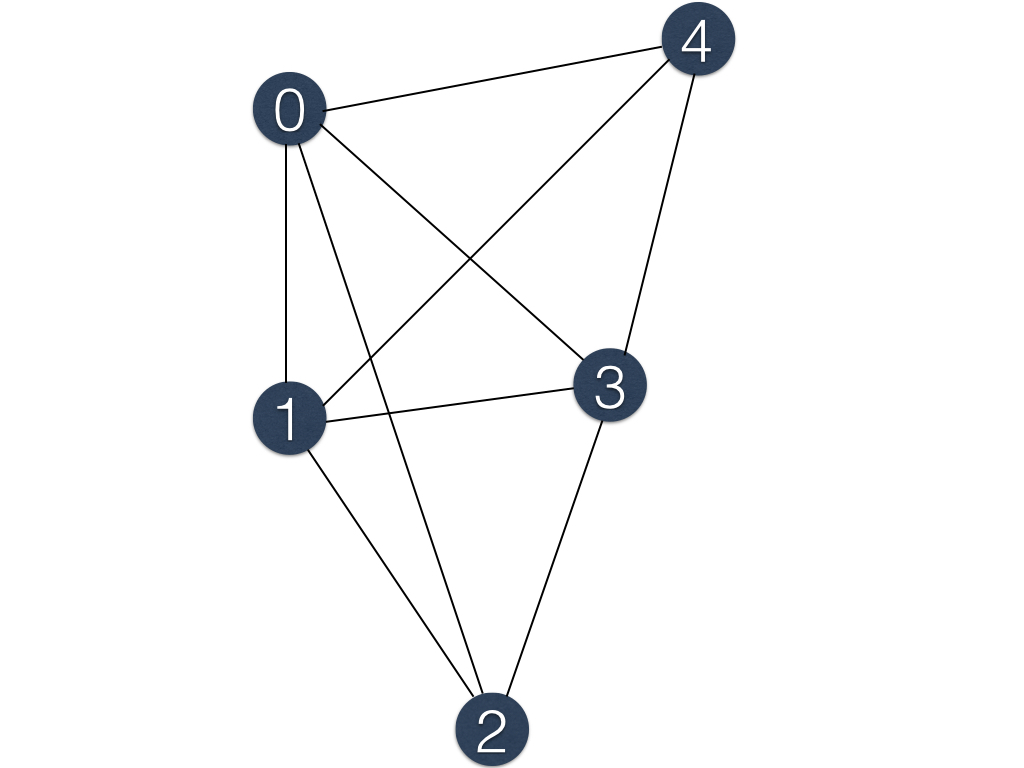}
        }
        \subfigure[3-truss]{
           \label{3_truss}
           \includegraphics[width=0.25\textwidth]{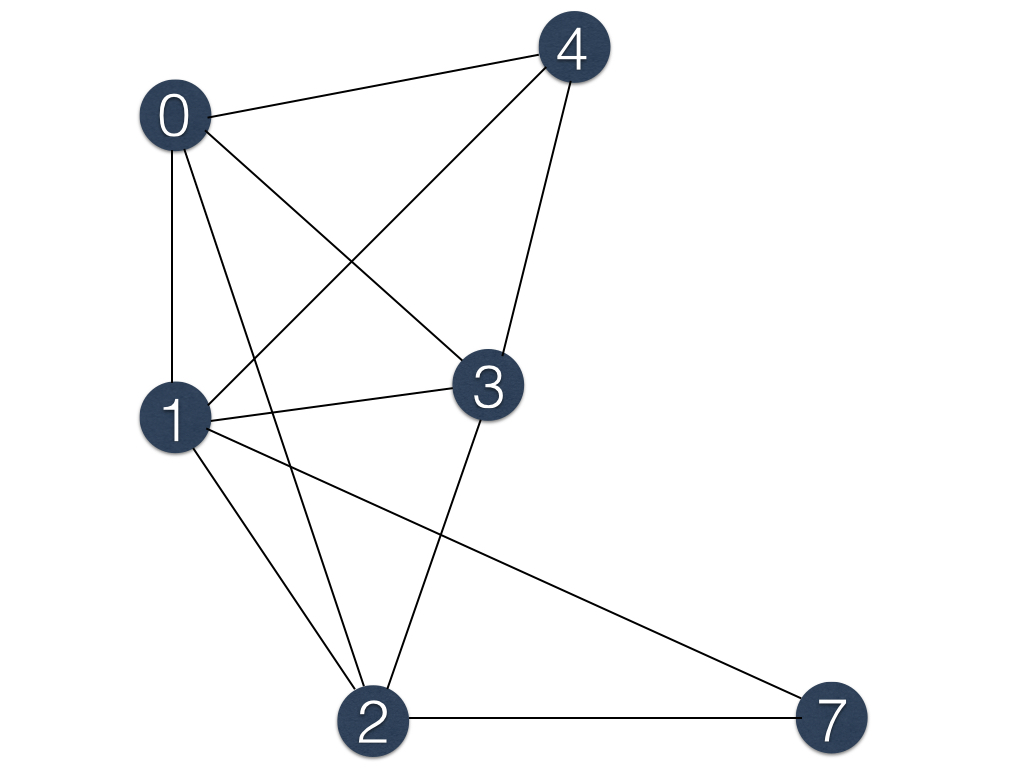}
        }
    \caption{
    Example using graph to represent crime event pairs/chains. 
     }
\label{event_graph}
\end{center}
\end{figure*}

Suppose we use a vertex $v$ to represent an event that occurred in location $(x, y)$ and at time $t$. Moreover, if two vertices $v_1$ and $v_2$ representing different events occurred within a given time and space constraint, an undirected edge $(v_1, v_2)$ will be used to connect them. If there are $V$ number of events and $E$ number of event pairs,
the resulting vertices and edges form a graph $G$ (In this article, we assume that there is only one undirected edge between any two vertices). If there exists a set of events with $n$ number of events, each event is paired with every other $n-1$ event. We call this set of events an event chain. A subgraph $g$ in $G$ can represent this event chain such that each vertex in the $g$ will have edges connecting every other vertex in $g$. Figure~\ref{origin} shows an example of $8$ crime events and $13$ event pairs. In the figure, subgraph induced from vertices $\{1, 2, 3\}$ shows an example of a 3-event chain which is a triangle, Furthermore, subgraph induced from vertices $\{0, 1, 3,4\}$ shows an example of a 4-event chain, which is a $4$-clique.
Press ENTER or type command to continue
The degree of a vertex $v$ is defined as the number of edges connecting $v$. Take Figure~\ref{origin} for example, the degree for vertex $1$ is $5$, and the degree for vertex $5$ is $2$. The definition of k-clique \cite{bomze1999maximum} is, each vertex in k-clique has degree of exactly $k-1$; Figure~\ref{3_clique} shows a 4-clique subgraph. Similarly, k-core \cite{batagelj2003m} is the subgraph in which each of its vertices has a degree of no less than $k$. Figure~\ref{3_core} shows a 3-core subgraph, k-DBSCAN \cite{birant2007st} is also a degree based cohesive subgraph, the method leverages k-degree vertices to greedily expand clusters(we will discuss the detail later), and Figure~\ref{origin} itself is the subgraph induced by 3-DBSCAN. As for an edge $e$ in $G$, the number of triangles it belongs to is called the support. For instance, the support for edge $(1,2)$ is $3$, and the support for edge $(2,7)$ is $1$. k-truss \cite{wang2012truss} is the subgraph with each of its edges having support no less than $k-2$; Figure~\ref{3_truss} shows a 3-truss induced from the original graph. The clustering coefficient \cite{watts1998collective} evaluates the tightness of the connection in a cohesive subgraph. If we use $coe(g)$ to denote a graph's clustering coefficient, empirically we have $coe(g_{k-clique}) \geq coe(g_{k-truss}) \geq coe(g_{k-core}) $ \cite{wang2012truss}.

\subsection{Near Repeat Event Chain Detection Algorithm}
\subsubsection{Algorithm description}
Given a set of crime events, we can represent each event using a coordinate $x$, $y$, and a time $t$ of crime type $p$. We would transform the coordinate using UTM format \cite{buchroithner2016geodetic}. The process of finding near repeat crime event chain can be formulated as the following two steps:

\begin{itemize}
\item Create a graph based on the spatial-temporal coordinates of a specific crime type; Since computing all pairs of events is expensive, we will build an R-tree \cite{guttman1984r} using 3-dimensional coordinates x, y, and t. A vertex forms edges with its neighbors by specifying some query criteria in R-tree.
\item Based on the graph created at step 1, compute the cohesive subgraphs such as k-clique, k-core, k-DBSCAN, or k-truss. Optimization methods might be applied; for instance, we can divide the graph into small graphs, if multiple connected components are detected \cite{cormen2009introduction}.
\end{itemize}

The algorithm is described in Figure~\ref{algo:event_chain}. The complexity of the algorithm could be divided into two steps. Suppose we have $V$ events, and $E$ event pairs. The complexity for the first graph generation process is dependent on data, and can not guarantee a worst-case complexity, but its lower bound is $O(V)$. The complexity of computing k-clique is NP-Hard \cite{bomze1999maximum}, k-core is $O(E)$ \cite{batagelj2003m}, k-DBSCAN is $O(Vlog(V))$ \cite{Gan2015DBSCAN}, and k-truss is $O(E^{1.5})$ \cite{wang2012truss}.

\begin{figure}[bp!]
\removelatexerror
\begin{algorithm}[H]
\SetKwData{Left}{left}\SetKwData{This}{this}\SetKwData{Up}{up}
\SetKwFunction{Union}{Union}\SetKwFunction{FindCompress}{FindCompress}
\SetKwInOut{Input}{input}\SetKwInOut{Output}{output}

    \KwIn{Set of crime events $C$, range criteria ($r_x$, $r_y$, $r_t$) }
    \KwOut{Set of near repeat event chains, $T$ }
    Initialize R-tree $R$ \;
    \For{$v$ in $C$}
    {
        $R.insert(v)$ \;
    }
    \For{$v$ in $C$}
    {
        ($x$, $y$, $t$) = coordinate of $v$ \;
        $S$ = R.retrieve([$x\pm r_x$, $y\pm r_y$, $t\pm r_t$]) \;
        \For{$u$ in $S$}
        {
            add edge ($u$, $v$) to $G$ \;
        }
    }
    $T$ = cohesive\_subgraph\_algo($G$) \;
    \Return{$T$} \;
\end{algorithm}
\caption{{ Near repeat event chain calculation.}}
\label{algo:event_chain}
\end{figure}

\begin{figure}[bp!]
\removelatexerror
\begin{algorithm}[H]
\SetKwData{Left}{left}\SetKwData{This}{this}\SetKwData{Up}{up}
\SetKwFunction{Union}{Union}\SetKwFunction{FindCompress}{FindCompress}
\SetKwInOut{Input}{input}\SetKwInOut{Output}{output}

\KwIn{ Graph $G$}
\KwOut{k-core $(k \geq 3)$ $T$ }
$k = 3$, $T_k=\emptyset$ \;
Compute deg($v$) for $v \in G$ \;
Sort all the vertices in ascending order by degree and place them in $U$ \;
\While{ $\exists v$ in $U$ such that $deg(v) < k$ }
{
    $W$ = $nb(v)$ \;
    \For{$w$ in $W$}
    {
        $deg(w) --$ \;
        $deg(w) --$ \;
        Reorder $w$ in $U$ according to its new degree;
    }
    Remove $v$ and its adjacent edges from $G$ \;
    Remove $v$ from $U$ \;
}
\If{ Not all $v$ in $U$ are removed}
{
    $T[k] = G$ \;
    $k ++$ \;
    goto step 4 \;
}
\Return{$T$} \;
\end{algorithm}
\caption{{k-core computation.}}
\label{algo:k_core}
\end{figure}

\begin{figure}[bp!]
\removelatexerror
\begin{algorithm}[H]
\SetKwData{Left}{left}\SetKwData{This}{this}\SetKwData{Up}{up}
\SetKwFunction{Union}{Union}\SetKwFunction{FindCompress}{FindCompress}
\SetKwInOut{Input}{input}\SetKwInOut{Output}{output}

\KwIn{ Graph $G$}
\KwOut{k-DBSCAN $(k \geq 3)$, $T$ }
$k = 3$, $T_k=\emptyset$ \;
compute deg($v$) for $v \in G$ \;
Sort all the vertices in descending order by degree and place them in $U$ \;

\While{ $\exists v$ such that $deg(v) \geq k$ and $visited(v) == false$ }
{
    expand($G$, $v$, $k$) \;
}

\While{ $\exists v$ such that $visited(v) == false$ }
{
    Remove $v$ and its adjacent edges from $G$ \;
    Remove $v$ from $U$ \;
}

\If{ Not all $v$ in $U$ are removed}
{
    $T[k] = G$ \;
    $k ++$ \;
    set all $v$ in $U$ as unvisited \;
    goto step 4 \;
}

\Return{$T$} \;
\end{algorithm}
\caption{{ k-DBSCAN computation.}}
\label{algo:k_dbscan}
\end{figure}

\begin{figure}[bp!]
\removelatexerror
\begin{algorithm}[H]
\SetKwData{Left}{left}\SetKwData{This}{this}\SetKwData{Up}{up}
\SetKwFunction{Union}{Union}\SetKwFunction{FindCompress}{FindCompress}
\SetKwInOut{Input}{input}\SetKwInOut{Output}{output}

\KwIn{ Graph $G$, vertex $v$, $k$}

$W$ = $nb(v)$ \;
\For{$w$ in $W$}
{
    \If{$visited(w) == false$}
    {
        $visited(w) = true$ \;
        \If{$deg(w) \geq k$}
        {
            expand(w) \;
        }
    }
}

\end{algorithm}
\caption{{Expand procedure.}}
\label{algo:expand}
\end{figure}

\begin{figure}[bp!]
\removelatexerror
\begin{algorithm}[H]
\SetKwData{Left}{left}\SetKwData{This}{this}\SetKwData{Up}{up}
\SetKwFunction{Union}{Union}\SetKwFunction{FindCompress}{FindCompress}
\SetKwInOut{Input}{input}\SetKwInOut{Output}{output}

\KwIn{ Graph $G$}
\KwOut{k-truss $(k \geq 3)$, $T$ }
$k = 3$, $T_k=\emptyset$ \;
compute sup($e$) for $e \in G$ \;
Sort all the edges in ascending order of their support and place them in $U$ \;
\While{ $\exists e$ in $U$ such that $sup(e) \leq (k-2)$ }
{
    $e$ = ($u$, $v$) with the lowest support \;
    $W$ = $nb(u) \cap nb(v)$ \;
    \For{$w$ in $W$}
    {
        $sup(u, w) --$ \;
        $sup(v, w) --$ \;
        Reorder (u, w) and (v, w) according to their new support;
    }
    Remove $e$ from $G$ \;
    Remove $e$ from $U$ \;
}
\If{ Not all $e$ in $U$ are removed}
{
    $T[k] = G$ \;
    $k ++$ \;
    goto step 4 \;
}
\Return{$T$} \;
\end{algorithm}
\caption{{ k-truss computation.}}
\label{algo:k_truss}
\end{figure}

\subsubsection{k-clique enumeration Revisited}
The maximum clique problem is a widely researched area, and there are lots of papers on this topic since the general algorithmic framework for clique enumeration algorithm is different from the other three algorithms, we will not spend too much effort on this theme.

\subsubsection{k-core Computation Revisited}
Figure~\ref{algo:k_core} displays the skeleton of the k-core algorithm. Vertices are sorted by their degrees in ascending order, and the criteria of $k$ start from 3. Vertices with degree less than $k$ and their adjacent edges are removed from the graph $G$ and the neighbor vertices of these removed vertices (we use $nb(v)$ to denote neighbors of $v$) will update their degrees accordingly. Once there is no such vertex to be removed, the remaining graph will be placed in the k-core class $T_k$, and $k$ will be incremented, and the removing procedure will start again. The procedure continues until there is no vertex to be removed.

\subsubsection{k-DBSCAN Computation Revisited}
k-DBSCAN is a density-based clustering algorithm. In this paper, we translate the k-DBSCAN algorithm into the equivalence of the cohesive subgraph algorithm. The algorithm is as Figure~\ref{algo:k_dbscan} shows. The algorithm finds a vertex $v$ with $deg(v) \geq k$ from $k=3$, then expand it (as Figure~\ref{algo:expand} shows), every expansion will result in the vertices in the expansion being marked as visited. If there is no vertex to be expanded, we will remove all the vertices that are not visited from $G$. The procedure continues until there is no vertex to be removed.

\subsubsection{k-truss Decomposition Revisited}
Truss decomposition is firstly introduced in paper \cite{cohen2008trusses} to detect possible subgroups within a social network. It is pretty useful in community detection. New and efficient algorithms are introduced to compute truss efficiently \cite{wang2012truss}. The idea of the algorithm is to compute the support for each edge first. For each edge, the $O(d)$ complexity algorithm for triangle enumeration will be applied, $d$ is the larger degree of the two vertices forming an edge. In this paper, we will use Compressed Sparse Row (CSR) \cite{saad1990basic} to store the edge array. The skeleton of the k-truss algorithm is shown in Figure~\ref{algo:k_truss}. Then the edges are sorted in the ascending order by their support. To compute the k-truss, every edge with support less than $k-2$, along with its incident vertices, will be removed. Moreover, the incident edges will update their support and their position in the edge array following similar methods k-core computation. Followed by the removal of all edges that do not form a k-truss, the remaining graph consists of k-trusses. Furthermore, the value of $k$ will be incremented, followed by the same edge removing steps until there are no edges left in the graph. In the algorithm, because the range of supports is already known, sorting can be done with $O(E)$ complexity using bucket sort.

\section{Experiments}

\subsection{Data Sets}
The data used in this research contains the real crime data onto New York (NYC), Washington DC (DC), and Chicago (CHI) retrieved from data.gov \cite{chidata}\cite{dcdata}\cite{nydata}. The general information about the data is displayed in TABLE~\ref{data_general}. In the table, the granularity of time is in a day(s), and the granularity of space is in meters. \#t means the number of crime types, \#d means the number of duplicated events. \#events are the number of events after combining duplications. We have removed the data that is not conformed to the right format (for instance, data that does not fall into the range in TABLE~\ref{data_general} ), and combined the duplicated entries ( for example, a crime of the same type happens at the same time of the same location, see TABLE~\ref{data_general} \#d). In general, all three data sets have crime numbers of a million scale.

Since there are numerous crime types of DC and CHI data (see TABLE~\ref{data_general} \#t), we only choose the crime types of burglary (BUR), robbery (ROB) and theft (TFT) for detailed discussion. We selected the spatial-temporal range limit $r_x=r_y=100 (meters)$ with $r_t=10 (days)$, and the feature of graphs generated applying this criteria have the property as TABLE~\ref{data_graph} shows. In the table, \#V is the number of vertices, \#E is the number of edges, \#CC is the number of connected components(we do not count the isolated vertices as CC), d\_avg and d\_var are the mean and variance of the diameters of connected components; c\_avg and c\_var are the mean and variance of clustering coefficient of connected components. We use Floyd Warshal \cite{cormen2009introduction} algorithm to calculate the all-pairs shortest path of each clique, and use this information to infer the diameter of each clique. As for the clustering coefficient \cite{watts1998collective}, we use it to evaluate how densely these graphs are organized. In general, burglary and robbery are sparse near repeat events in comparison to theft concerning the number of vertices $\#V$, which is also indicated by a more substantial number of edges and connected components $\#CC and \#E$. The diameter and clustering coefficient feature also indicates that theft has large clusters, and these clusters are dense. Inferred from the table, the graphs in all data obey small-world property because the diameters of the graphs are small \cite{watts1998collective}.

\begin{table}[htbp]
\centering
\caption{GENERAL INFORMATION OF GRAPHS.} 
\label{data_graph}
\begin{tabular}{llllllll}
\toprule
    & \#V    & \#E      & \#CC  & d\_avg & d\_var & c\_avg & c\_var \\ \midrule
\multicolumn{8}{c}{NY}                                                        \\ \midrule 
BUR & 187k & 112k   & 24k & 1.25      & 0.64      & 0.12         & 0.064      \\ 
ROB & 198k & 152k   & 27k & 1.34      & 1.38      & 0.13         & 0.068      \\ 
TFT & 421k & 1.5m   & 55k & 1.77      & 6.66      & 0.20         & 0.089      \\ \midrule 
\multicolumn{8}{c}{DC}                                                        \\ \midrule
BUR & 156k & 54k    & 13k & 1.26      & 0.80      & 0.10         & 0.054      \\ 
ROB & 54k  & 32k    & 6k  & 1.40      & 1.30      & 0.138        & 0.069      \\ 
TFT & 344k & 1.1m   & 33k & 1.75      & 7.78      & 0.17         & 0.080      \\ \midrule
\multicolumn{8}{c}{CHI}                                                       \\ \midrule
BUR & 197k & 118k   & 29k & 1.29      & 0.56      & 0.12         & 0.060      \\ 
ROB & 124k & 68k    & 14k & 1.34      & 0.96      & 0.11         & 0.058      \\ 
TFT & 650k & 3.4m   & 89k & 1.84      & 7.95      & 0.19         & 0.081      \\ \bottomrule 
\end{tabular}
\end{table}

\begin{table*}[htbp]
\centering
\caption{GENERAL INFORMATION OF DATA.}
\label{data_general}
\small
\begin{tabular}{llllllllll} \toprule
name & earliest & latest   & min x   & max x    & min y   & max y   & \#events & \#t      & \#d      \\ \midrule
NY   & 2006/06/04 & 2015/12/31 & 134239  & 1067186  & 121080  & 7220451 & 1123221    & 7        & 29k    \\ \midrule
DC   & 1978/01/01 & 2015/12/31 & 4840550 & 18915876 & 777144  & 8480189 & 2130867    & 43       & 89k    \\ \midrule
CHI  & 2001/01/01 & 2015/12/31 & 1092706 & 1205119  & 1813894 & 1951610 & 3102758    & 35       & 54k    \\ \bottomrule 
\end{tabular}
\end{table*}

\subsection{Knox test with Map-reduce}
Firstly, we prove the existence of a near-repeat effect in a real big data set by conducting a Knox test on the data set.  We implement the Knox test using the Map-reduce framework on Amazon AWS EMR and store the input/output on S3. The program is written in python and runs with Hadoop streaming \cite{White2010Hadoop} mode. For New York and Chicago theft data set, we used a cluster of 16 nodes,  and for all the other data sets, we used a cluster of 4 nodes (for a budget reason).

The computational time is recorded and is shown on TABLE~\ref{knox_time}. To the best of our knowledge, the previous Knox test research on crime data is orders of magnitudes smaller than our data. Since the complexity of the Knox test is $O(n^2)$, it is not practical to compare the timing of these results against the previous experiments. Hence in this paper, we claim that our method can finish the Knox test within a reasonable time from less than an hour to approximately 10 hours using big real-world data, which is not possible with the previous methods.

\begin{figure*}[htbp]
    \begin{center}
	\subfigure[NY BUR]{
            \label{NY_BUR_EMR}
            \includegraphics[width=0.31\textwidth]{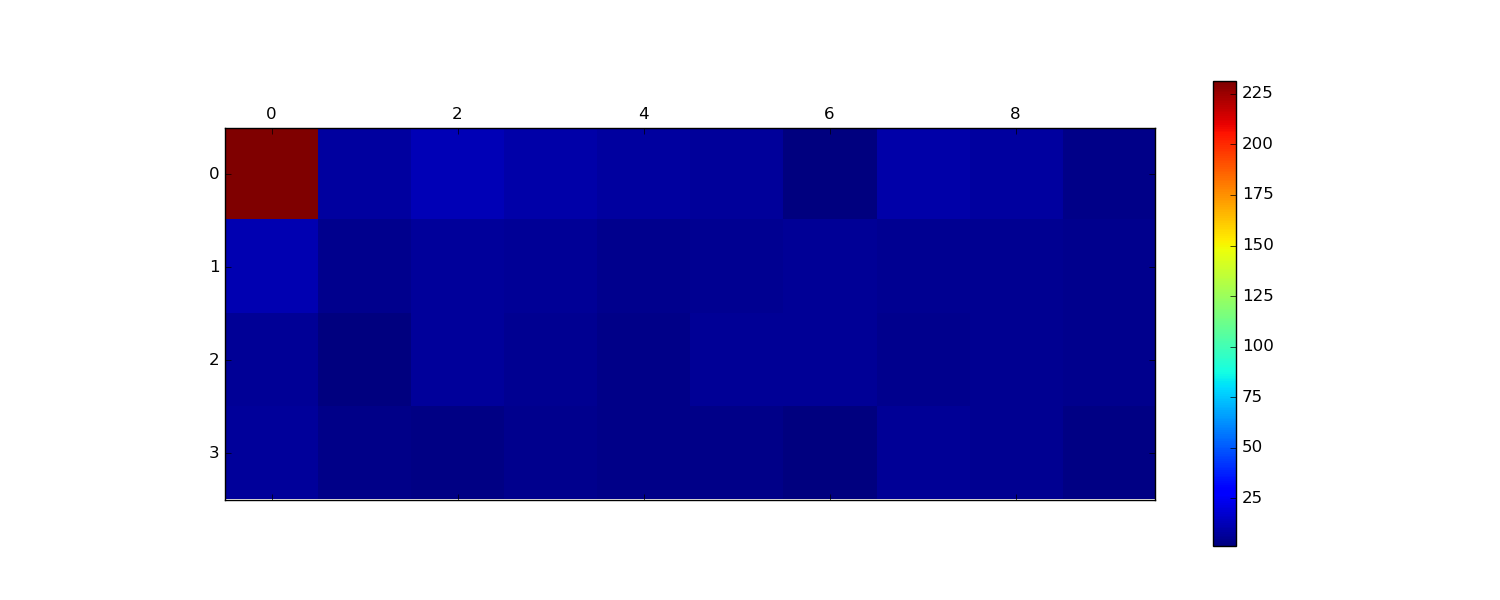}
        }%
        \subfigure[DC BUR]{
           \label{DC_BUR_EMR}
           \includegraphics[width=0.31\textwidth]{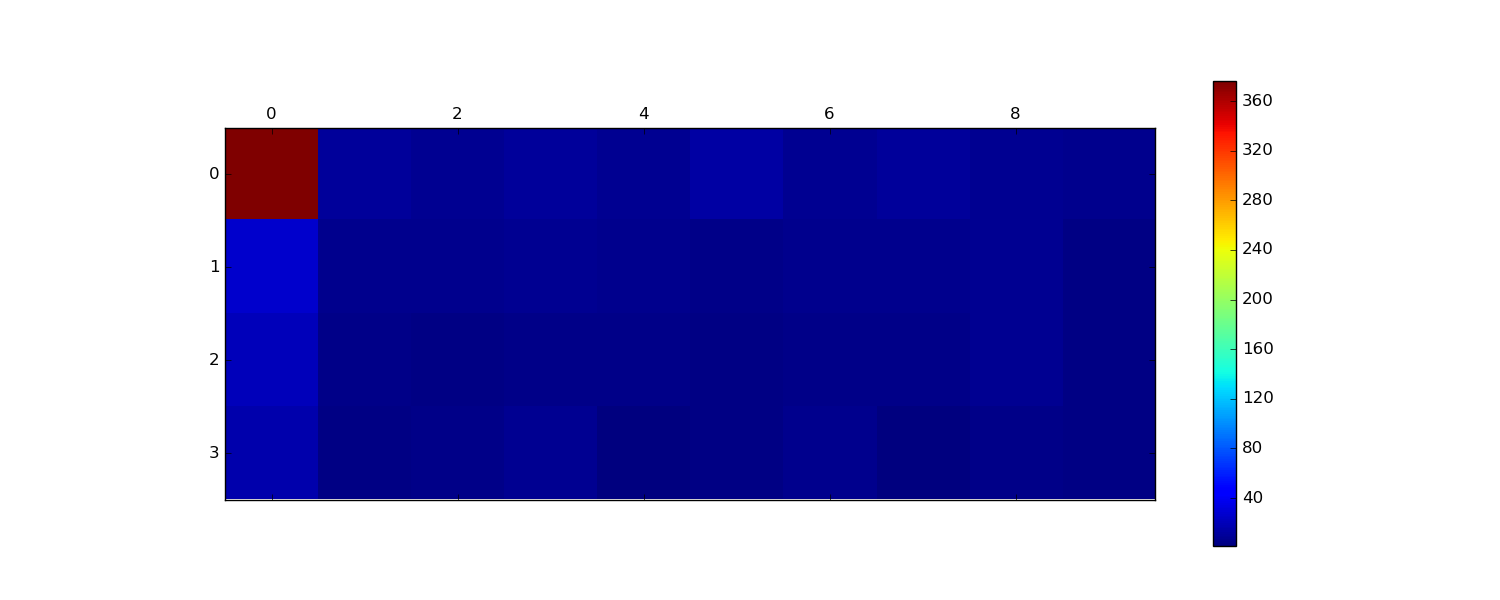}
        }
        \subfigure[CHI BUR]{
           \label{CHI_BUR_EMR}
           \includegraphics[width=0.31\textwidth]{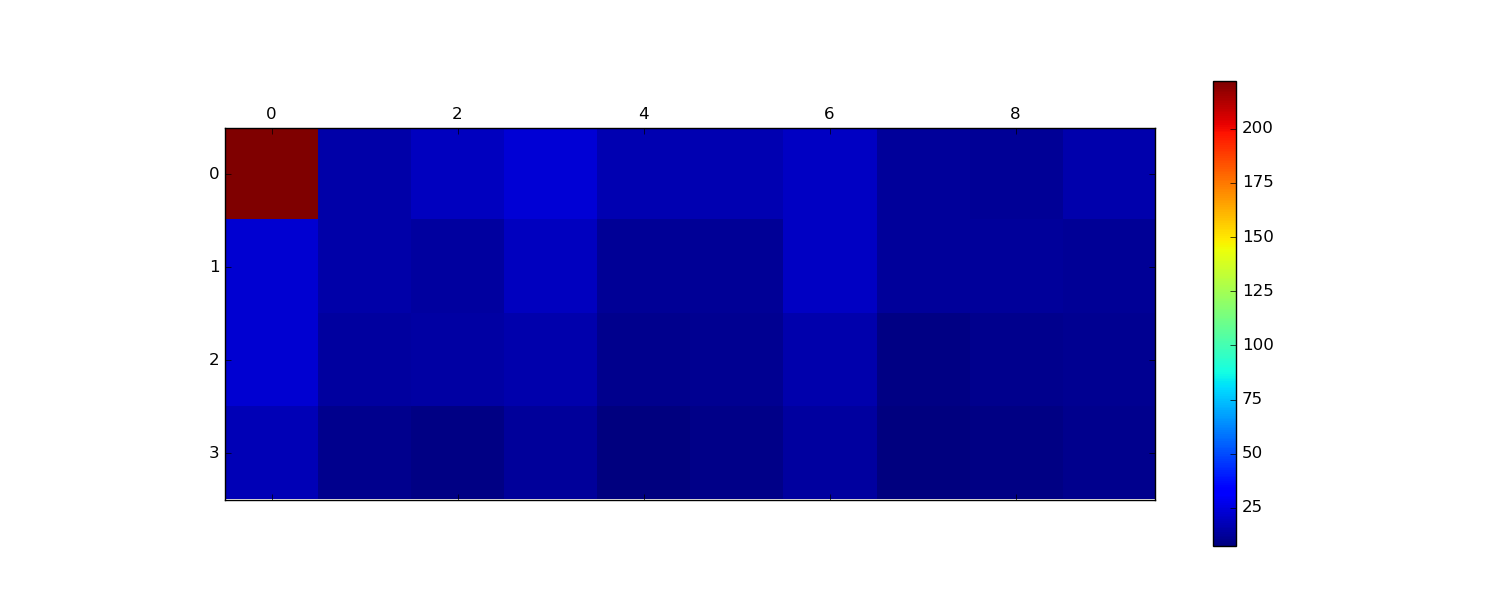}
        }
	\subfigure[NY ROB]{
            \label{NY_ROB_EMR}
            \includegraphics[width=0.31\textwidth]{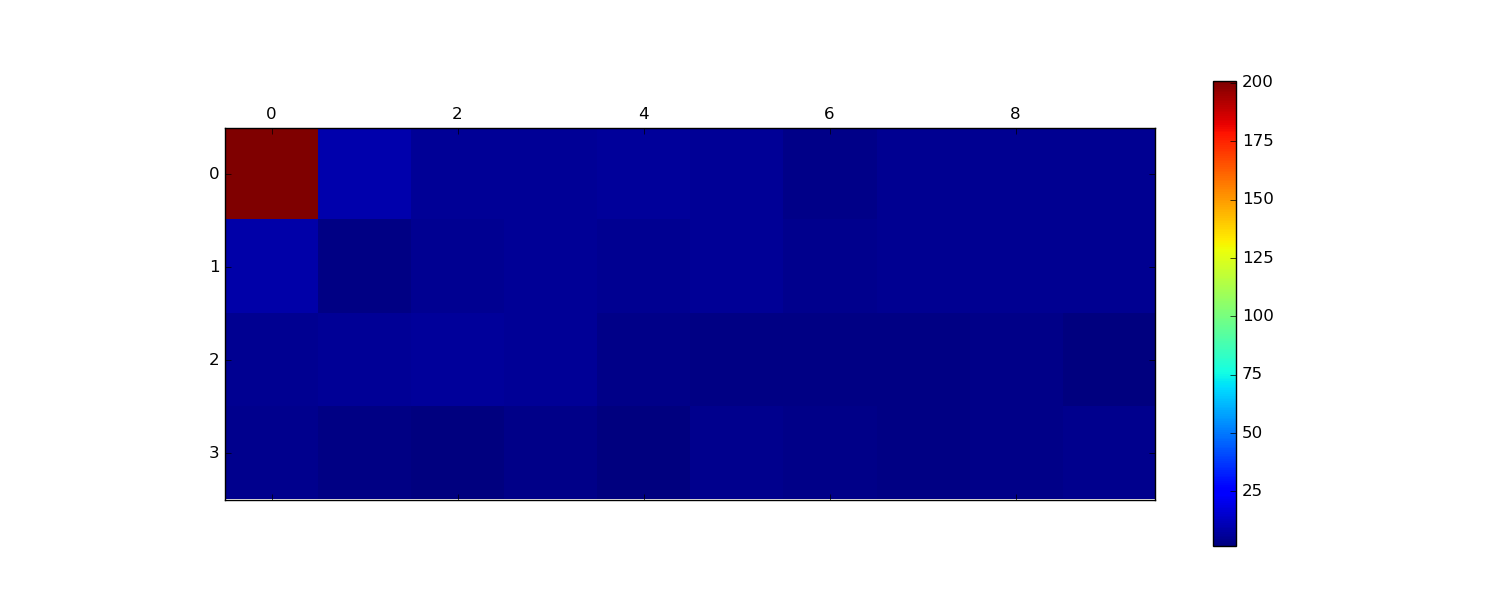}
        }%
        \subfigure[DC ROB]{
           \label{DC_ROB_EMR}
           \includegraphics[width=0.31\textwidth]{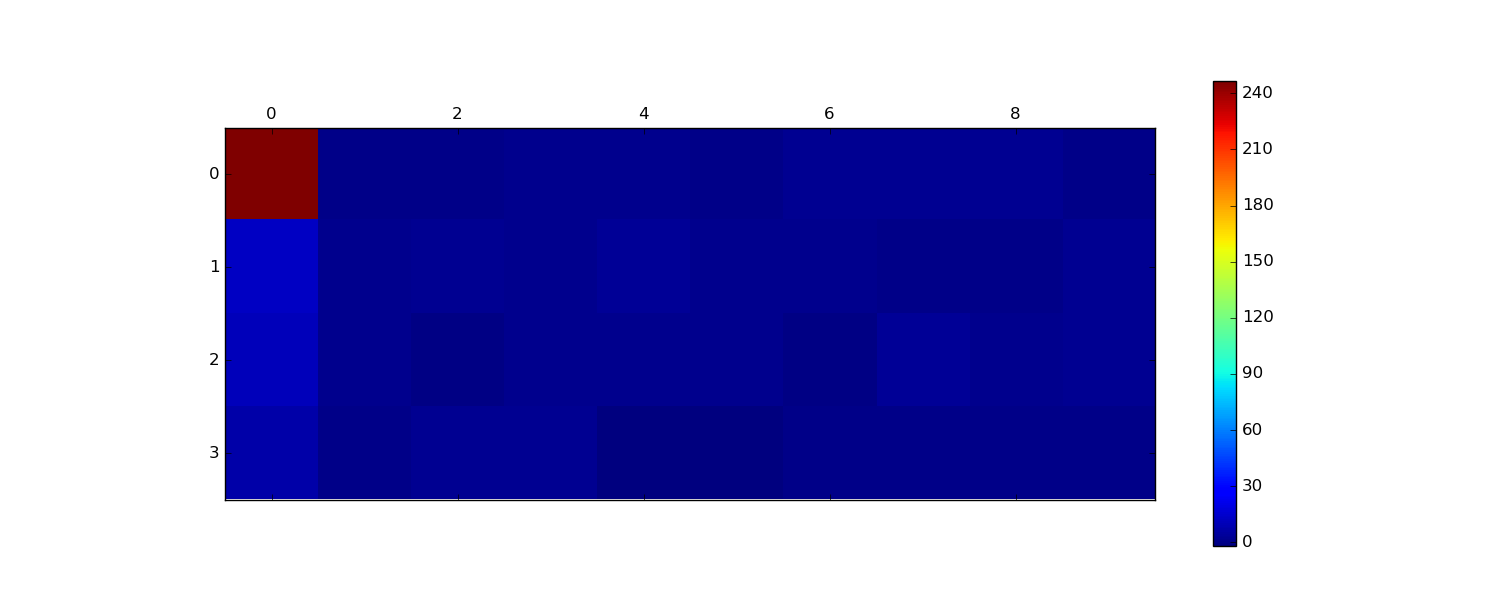}
        }
        \subfigure[CHI ROB]{
           \label{CHI_ROB_EMR}
           \includegraphics[width=0.31\textwidth]{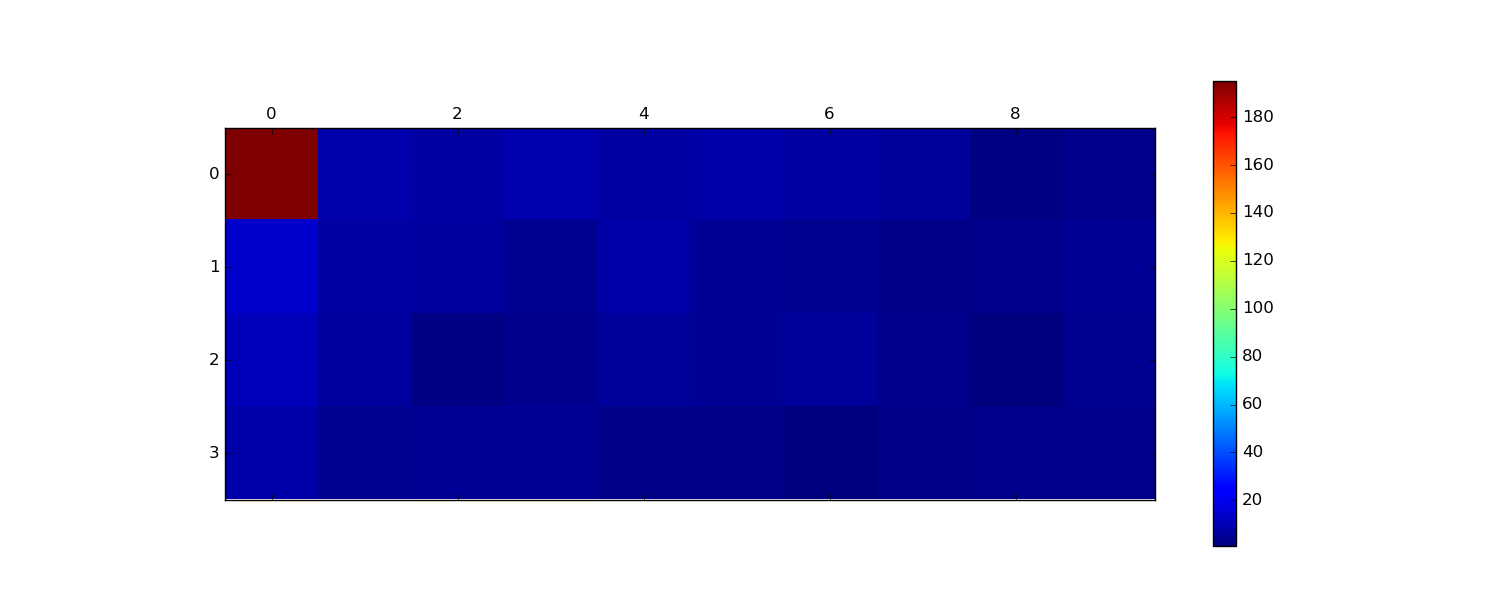}
        }
	\subfigure[NY TFT]{
            \label{NY_TFT_EMR}
            \includegraphics[width=0.31\textwidth]{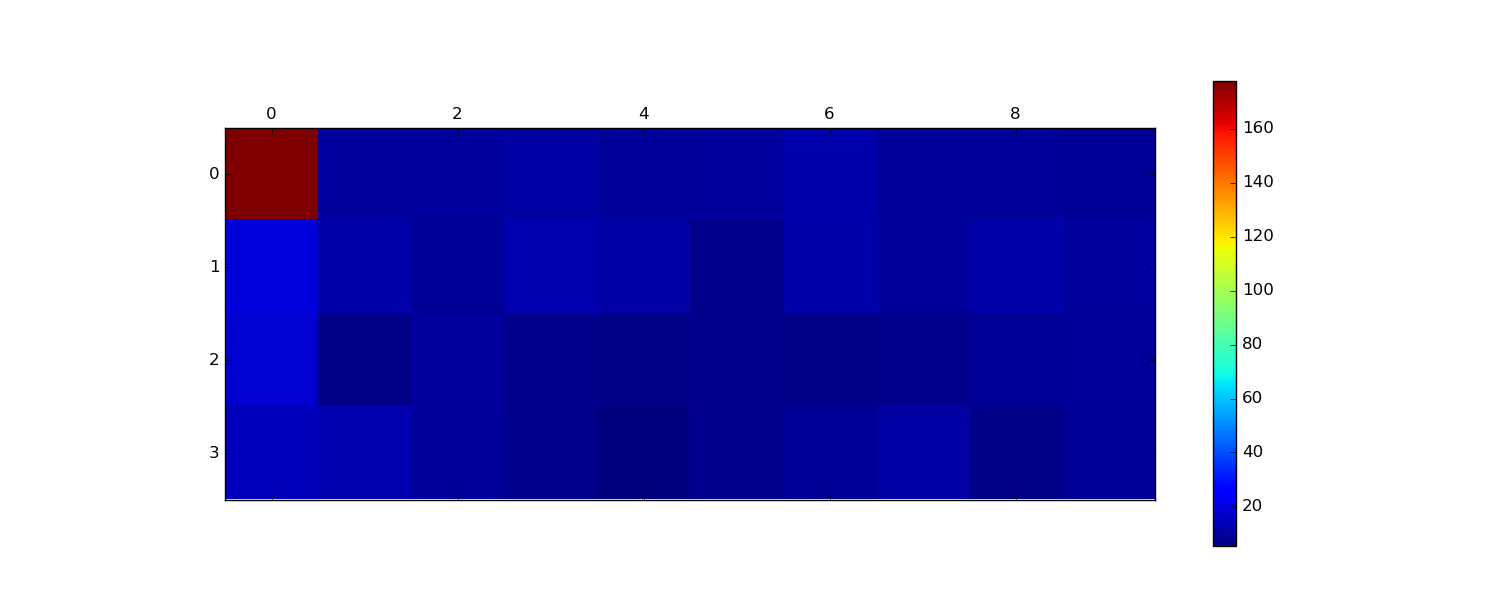}
        }%
        \subfigure[DC TFT]{
           \label{DC_TFT_EMR}
           \includegraphics[width=0.31\textwidth]{./DC_BUR_emr.png}
        }
        \subfigure[CHI TFT]{
           \label{CHI_TFT_EMR}
           \includegraphics[width=0.31\textwidth]{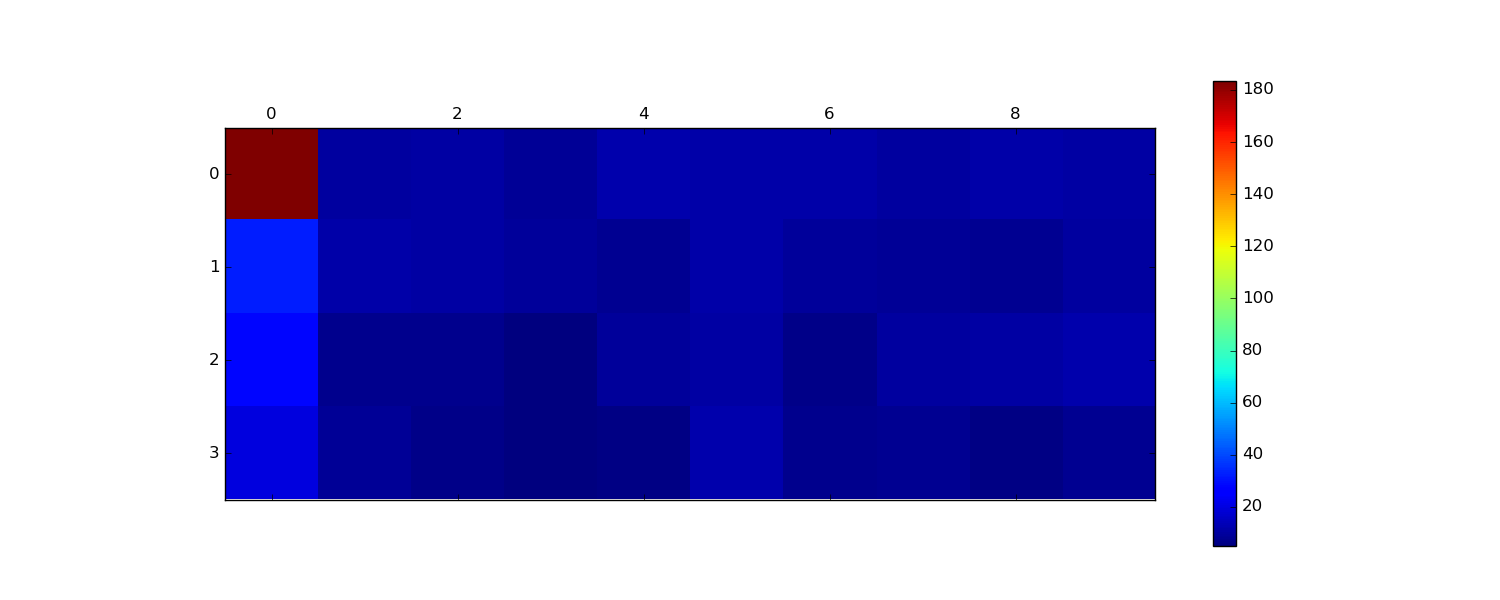}
        }
     \end{center}
    \caption{
        Knox test results. A $4 \times 10$ colored matrices represent each test, the row step is 14 days, and the column step is 100 meters. The residual value of that range plots the color of the entry.
    }
    \label{matshow}
\end{figure*}

To construct the Knox test table \cite{townsley2003infectious}, we have set the distance step as 100 meters and the time step as 14 days. The Knox test result is shown in Figure~\ref{matshow} using a heatmap. In the figure, we do not show the result of distance larger than $10 \times 100=1000$ meters and time difference larger than $4 \times 14=56$ days. It is evident from the heatmap that all three crime types in three cities exhibit near repeat effects because the upper left corner entries of each Knox test matrices have residual values that are significantly larger than other entries.

\begin{table}[htbp]
\centering
\caption{COMPUTATIONAL TIME (IN SECONDS) FOR KNOX TEST USING MAPREDUCE.}
\label{knox_time} 
\begin{tabular}{llll} \toprule
    City & BUR             & ROB             & TFT              \\ \midrule  
    NY   & 13320 (4 nodes) & 13980 (4 nodes) & 13560 (16 nodes) \\ 
    DC   & 9240 (4 nodes)  & 1440 (4 nodes)  & 34320 (4 nodes)  \\ 
    CHI  & 14340 (4 nodes) & 6000 (4 nodes)  & 24060 (16 nodes) \\ \bottomrule 
\end{tabular}
\end{table}

\subsection{Near-repeat chain detection}
We implement the k-core, k-DBSCAN and k-truss algorithm using C++, and GCC compiler with the c++-11 features enabled; Furthermore, the code is freely available on GitHub with the package name OPTKIT. To build the spatial-temporal index with an R-tree package, we use the open-source implementation of \cite{guttman1984r}. As for the k-clique, and the graph properties, we use the boost graph library (BGL) \cite{siek2001boost}. The experiment was run on an AWS machine \cite{amazon2015amazon}, with an m4.4xlarge Redhat instance. The instance has $16$ cores, and each has a 2.4 GHz Intel Xeon® E5-2676 v3 (Haswell) processor and $64$ Gbs of memory, the disk size is $160$ Gbs of EBS storage; the operating system is RHEL-7.2. Computational time is recorded by analyzing the log result using the glog library.

\begin{figure*}[htbp]
    \begin{center}
	\subfigure[NY BUR]{
            \label{NY_BUR_COE}
            \includegraphics[width=0.31\textwidth]{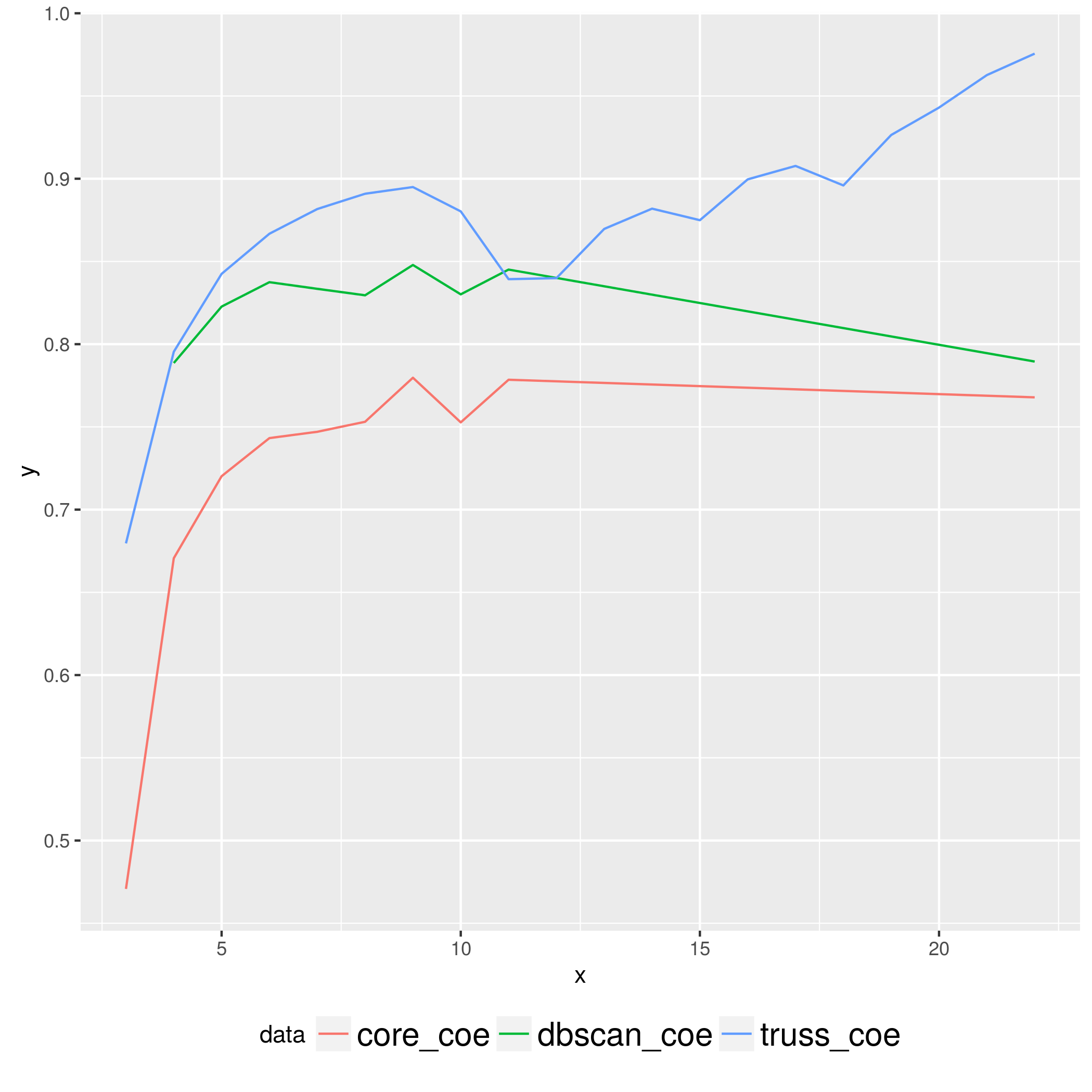}
        }%
        \subfigure[DC BUR]{
           \label{DC_BUR_COE}
           \includegraphics[width=0.31\textwidth]{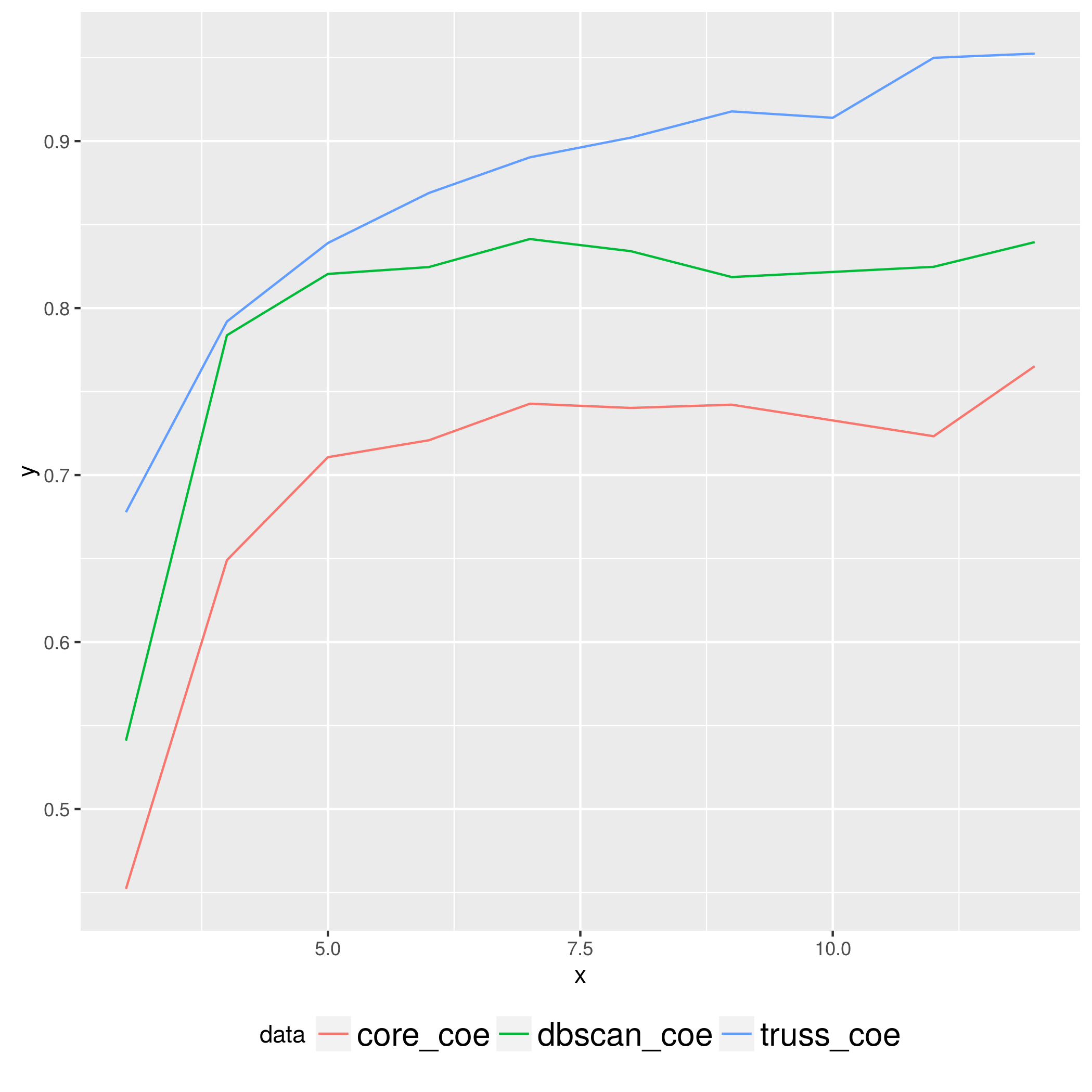}
        }
        \subfigure[CHI BUR]{
           \label{CHI_BUR_COE}
           \includegraphics[width=0.31\textwidth]{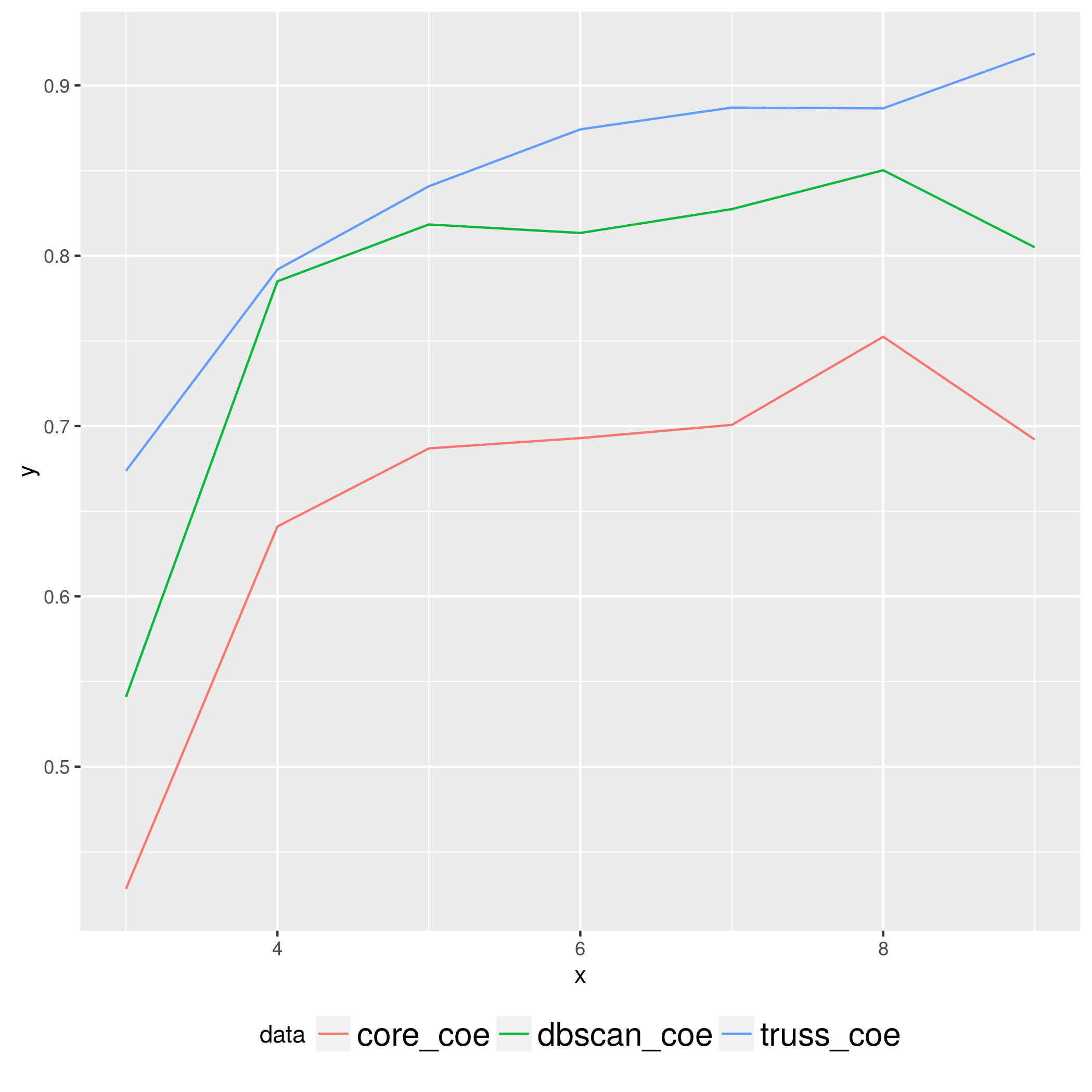}
        }
	\subfigure[NY ROB]{
            \label{NY_ROB_COE}
            \includegraphics[width=0.31\textwidth]{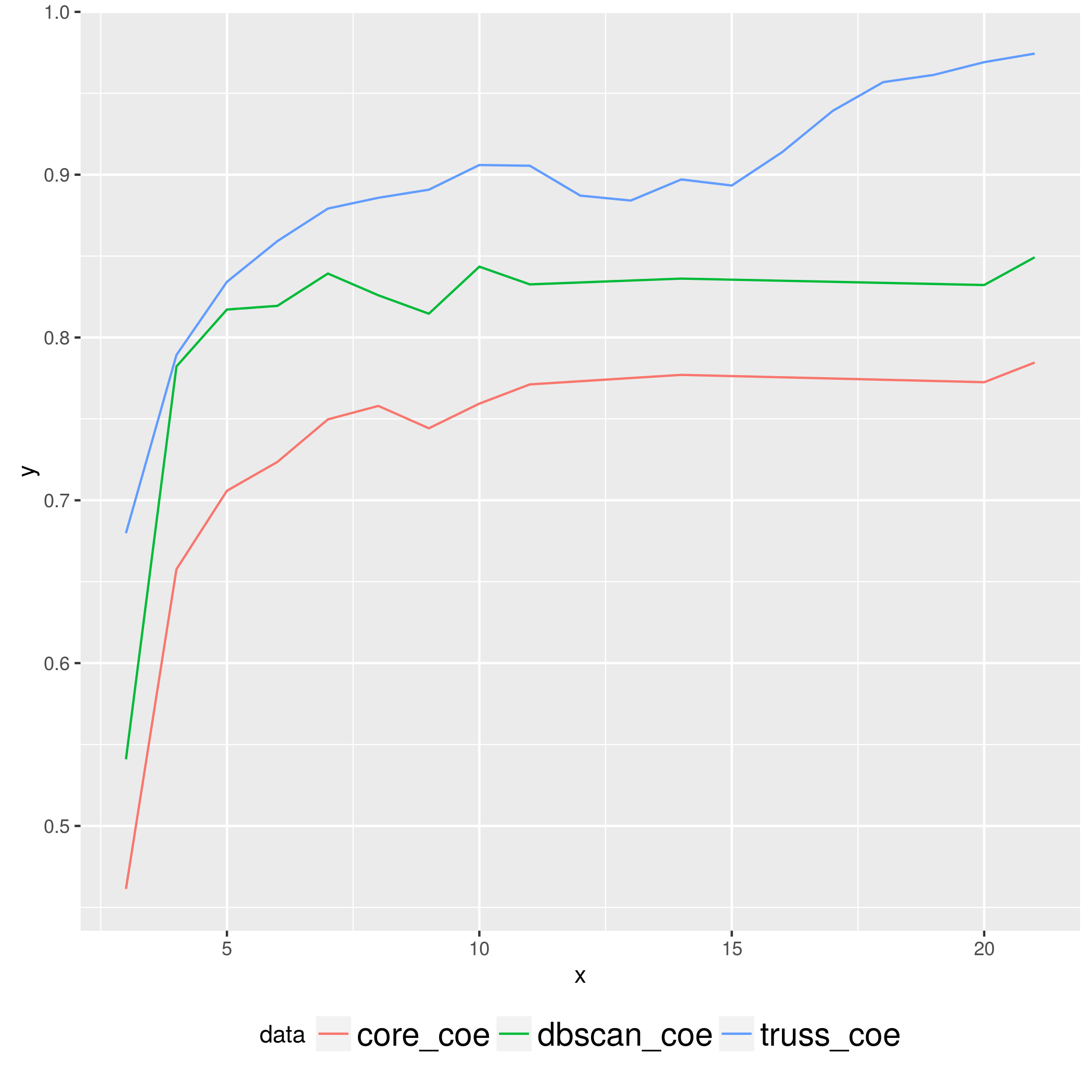}
        }%
        \subfigure[DC ROB]{
           \label{DC_ROB_COE}
           \includegraphics[width=0.31\textwidth]{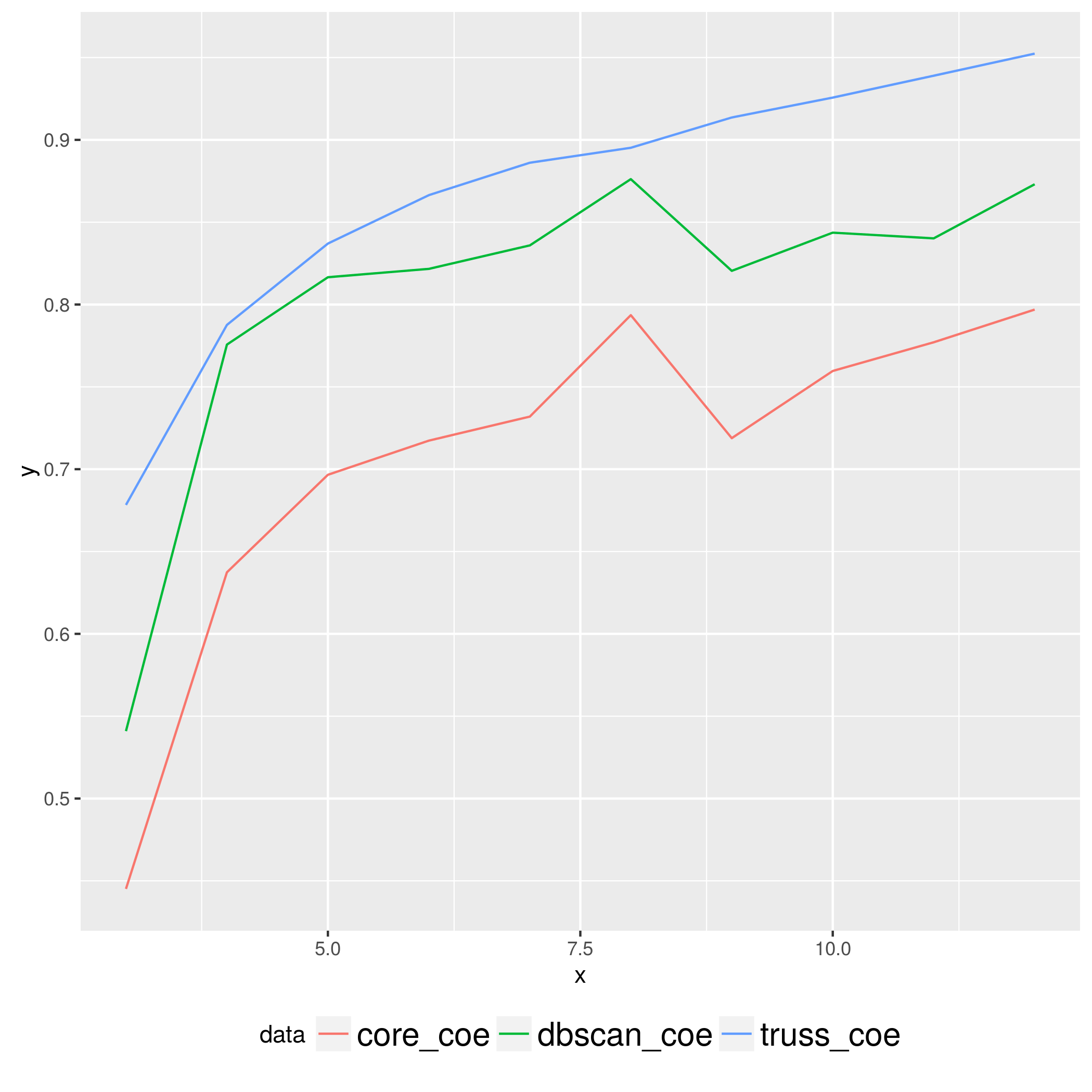}
        }
        \subfigure[CHI ROB]{
           \label{CHI_ROB_COE}
           \includegraphics[width=0.31\textwidth]{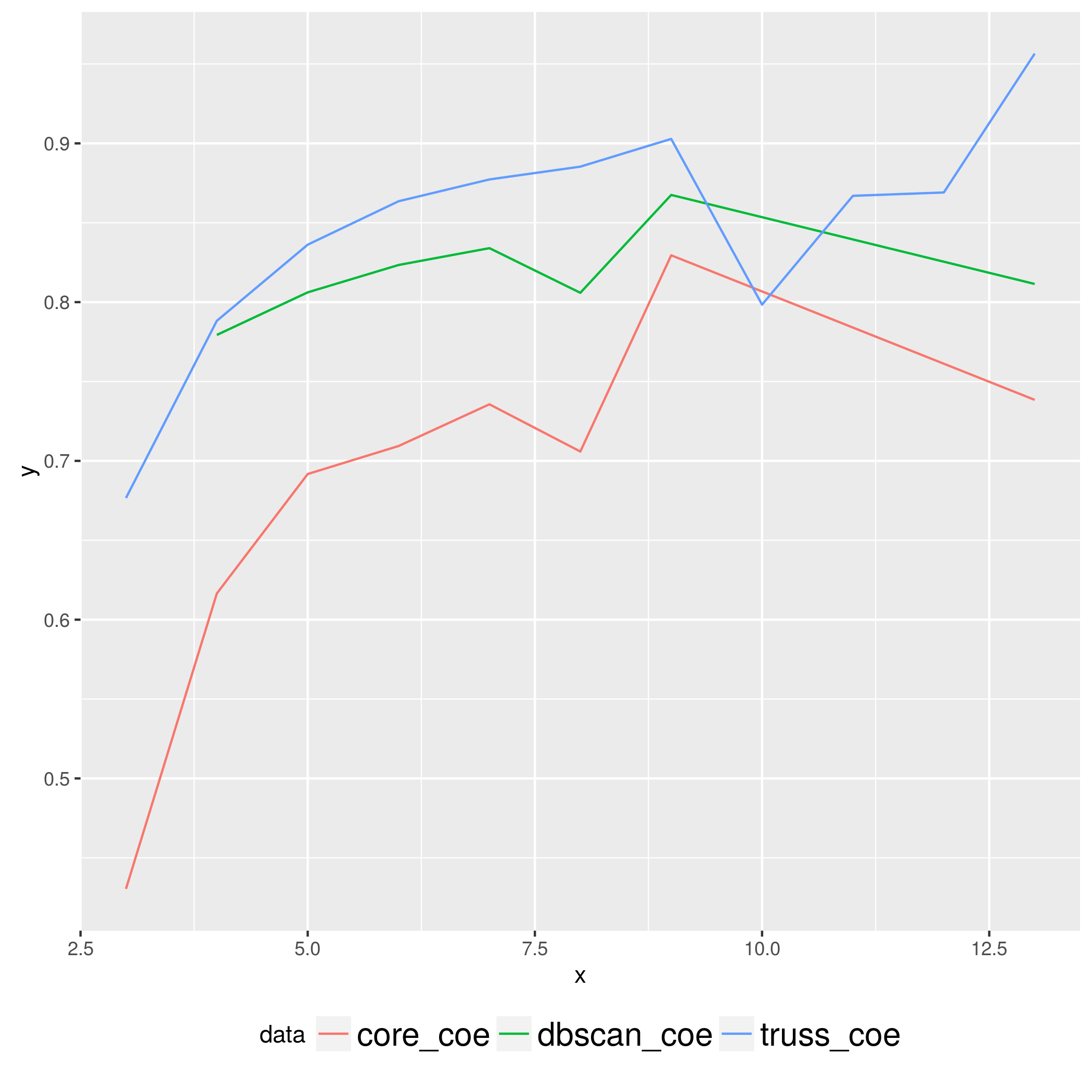}
        }
	\subfigure[NY TFT]{
            \label{NY_TFT_COE}
            \includegraphics[width=0.31\textwidth]{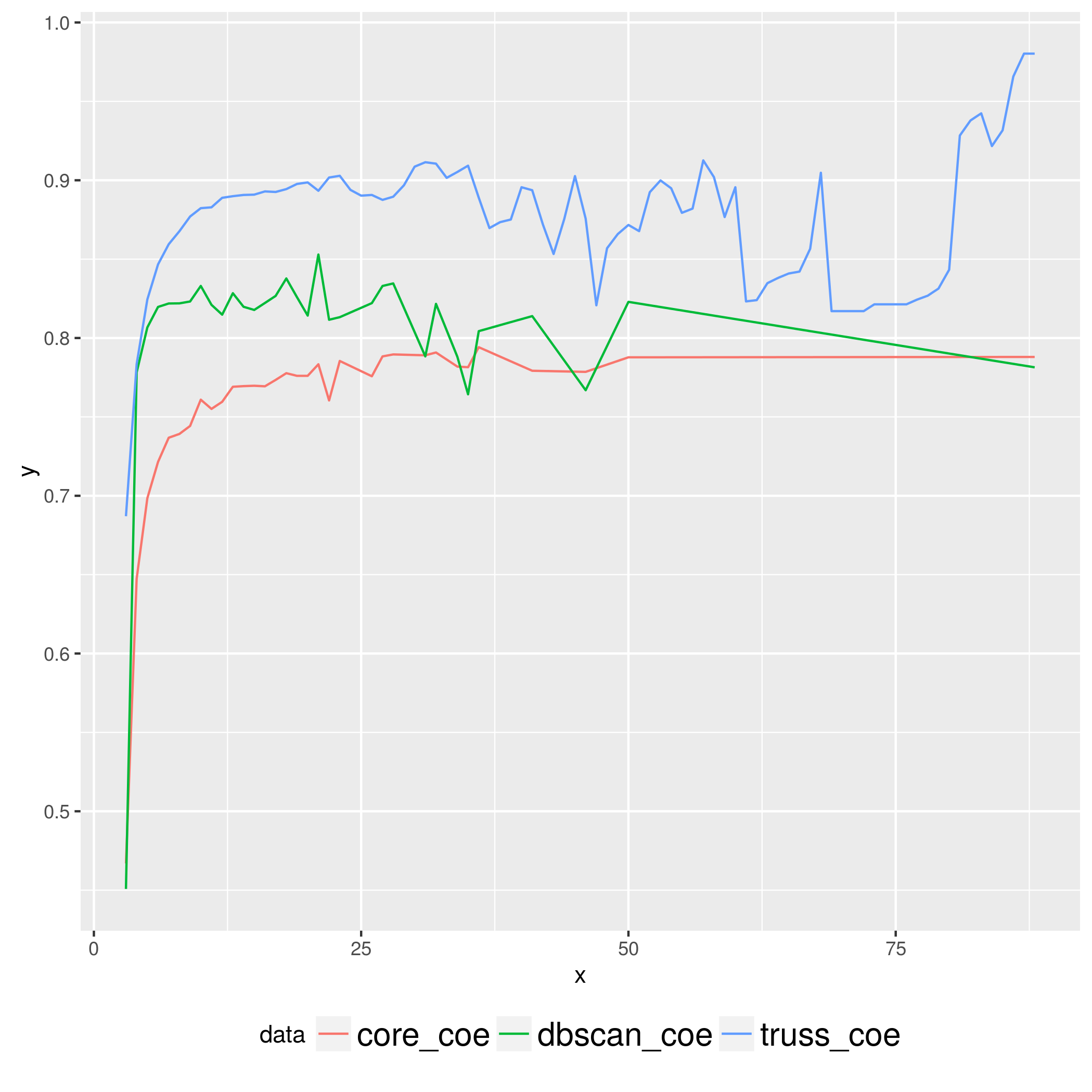}
        }%
        \subfigure[DC TFT]{
           \label{DC_TFT_COE}
           \includegraphics[width=0.31\textwidth]{./DC_BUR_coe.png}
        }
        \subfigure[CHI TFT]{
           \label{CHI_TFT_COE}
           \includegraphics[width=0.31\textwidth]{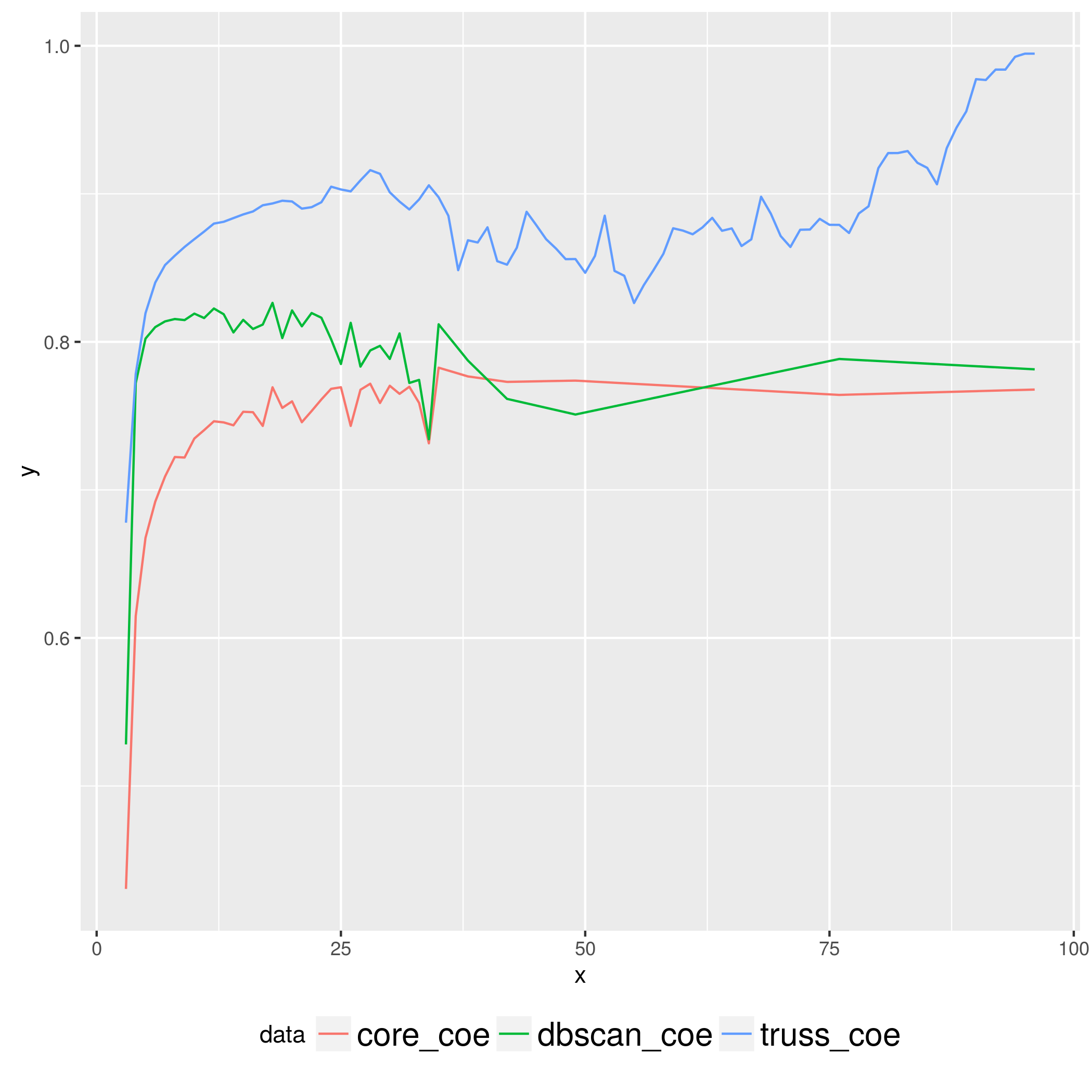}
        }
    \end{center}
        \caption{
        The clustering coefficient of cohesive subgraphs detected on different data (the x-axis is $k$, and the y-axis is the cluster coefficient value).
     }
    \label{coeshow}
\end{figure*}

\subsubsection{Computational Time}

The computational time is divided into seven parts,  including the time to 1) load the data,  which includes parsing and reading spatial-temporal coordinates of CSV format.  2) build R-tree and edges based on querying the R-tree.  3) separate edges based on the connected component computation using BGL. 4-6) implement the k-truss, k-core, and k-DBSCAN algorithms.  7) calculate k-cliques.

\begin{table}[htbp]
\centering
\caption{RESULTS FOR COMPUTATIONAL TIME.} 
\label{table_time}
\scriptsize
\begin{tabular}{lllllllll}
\toprule
    & load & R-tree & edges  & CC     & truss & core & dbscan & BGL    \\ \midrule
\multicolumn{7}{c}{NY}                                  \\ \midrule
BUR & 0.54 & 0.50   & 0.15   & 0.11   & 6.15  & 1.53 & 4.30 & 2.14   \\ 
ROB & 0.62 & 0.51   & 0.21   & 0.13   & 7.05  & 1.97 & 4.16 & 2.00    \\ 
TFT & 1.24 & 1.10   & 1.06   & 0.80   & 10.55 & 4.05 & 4.71 & 302.79 \\ \midrule
\multicolumn{7}{c}{DC}                                  \\ \midrule
BUR & 0.47 & 0.43   & 0.17   & 0.05   & 4.97  & 0.91 & 1.59 & 0.83   \\ 
ROB & 0.15 & 0.13   & 0.05   & 0.03   & 0.86  & 0.28 & 0.32 & 0.44   \\ 
TFT & 1.07 & 0.96   & 0.59   & 0.44   & 7.87  & 1.87 & 2.37 & 87.31  \\ \midrule 
\multicolumn{7}{c}{CHI}                                 \\ \midrule
BUR & 0.62 & 0.54   & 0.21   & 0.12   & 10.08 & 1.65 & 2.6  & 1.67   \\ 
ROB & 0.39 & 0.33   & 0.13   & 0.07   & 6.09  & 1.06 & 2.11 & 0.97   \\ 
TFT & 1.10 & 1.69   & 4.33   & 2.76   & 21.20 & 7.82 & 10.93& 487.92 \\ \bottomrule
\end{tabular}
\end{table}

\begin{table}[htbp]
\centering
\caption{NUMBER OF K-CLIQUE DETECTED.}
\label{table_clique}
\scriptsize
\begin{tabular}{lllllllll}
\toprule
    & 3 & 4 & 5  & 6 & 7 & 8 & 9 & $\geq$ 10 \\ \midrule
\multicolumn{7}{c}{NY}                                  \\ \midrule
BUR  & 4117  & 935  & 304  & 87  & 33  & 11  & 5  & 16 \\ 
ROB  & 4867  & 1364  & 491  & 195  & 88  & 46  & 24  & 40    \\ \midrule 
\multicolumn{7}{c}{DC}                                  \\ \midrule
BUR  & 1770  & 432  & 133  & 56  & 26  & 11  & 6  & 4 \\ 
ROB  & 1073  & 314  & 118  & 45  & 20  & 12  & 8  & 8 \\ \midrule
\multicolumn{7}{c}{CHI}                                 \\ \midrule
BUR  & 4941  & 1039  & 223  & 41  & 10  & 5  & 1 & 0\\ 
ROB  & 2338  & 600  & 207  & 66  & 34  & 12  & 2  & 4 \\ \bottomrule
\end{tabular}
\end{table}

\begin{table}[htbp]
\centering
\caption{NUMBER OF K-CORES DETECTED.}
\label{table_core}
\scriptsize
\begin{tabular}{lllllllll}
\toprule
    & 3 & 4 & 5  & 6 & 7 & 8 & 9 & $\geq$ 10 \\ \midrule
\multicolumn{7}{c}{NY}                                  \\ \midrule
BUR  & 1336  & 1295  & 803  & 307  & 163  & 60  & 35  & 94    \\ 
ROB   & 1732  & 1944  & 1144  & 624  & 351  & 233  & 177  & 321  \\ 
TFT  & 5227  & 7467  & 5787  & 4452  & 3124  & 2272  & 1824  & 8661  \\ \midrule
\multicolumn{7}{c}{DC}                                  \\ \midrule
BUR  & 641  & 664  & 306  & 193  & 114  & 66  & 52  & 34   \\ 
ROB   & 415  & 496  & 285  & 159  & 87  & 42  & 33  & 68  \\ 
TFT    & 2980  & 4090  & 2934  & 2415  & 1490  & 1153  & 951  & 6855 \\ \midrule 
\multicolumn{7}{c}{CHI}                                 \\ \midrule
BUR  & 2092  & 1729  & 649  & 195  & 40  & 33  & 13 & 0\\ 
ROB   & 481  & 315  & 153  & 105  & 62  & 6  & 18  & 0\\ 
TFT  & 11223  & 12904  & 8549  & 6081  & 4329  & 3047  & 2234  & 15028 \\ \bottomrule
\end{tabular}
\end{table}

\begin{table}[htbp]
\centering
\caption{NNUMBER OF K-DBSCAN DETECTED.}
\label{table_dbscan}
\scriptsize
\begin{tabular}{lllllllll}
\toprule
    & 3 & 4 & 5  & 6 & 7 & 8 & 9 & $\geq$ 10 \\ \midrule
\multicolumn{7}{c}{NY}                                  \\ \midrule
BUR   & 782  & 509  & 211  & 101  & 41  & 24  & 47  & 0\\ 
ROB    & 1  & 1068  & 707  & 389  & 243  & 148  & 110  & 224  \\ 
TFT      & 1  & 3894  & 3391  & 2744  & 1967  & 1473  & 1187  & 5193 \\ \midrule
\multicolumn{7}{c}{DC}                                  \\ \midrule
BUR     & 1  & 362  & 194  & 124  & 77  & 45  & 31  & 29  \\ 
ROB   & 415  & 496  & 285  & 159  & 87  & 42  & 33  & 68  \\ 
TFT   & 2980  & 4090  & 2934  & 2415  & 1490  & 1153  & 951  & 6855  \\ \midrule 
\multicolumn{7}{c}{CHI}                                 \\ \midrule
BUR  & 6  & 944  & 396  & 112  & 24  & 20  & 7  & 0\\ 
ROB  & 481  & 315  & 153  & 105  & 62  & 6  & 18  & 0\\ 
TFT    & 57  & 6071  & 4676  & 3547  & 2630  & 1859  & 1400  & 8608  \\ \bottomrule
\end{tabular}
\end{table}

\begin{table}[htbp]
\centering
\caption{NUMBER OF K-TRUSS DETECTED.}
\label{table_truss}
\scriptsize
\begin{tabular}{lllllllll}
\toprule
    & 3 & 4 & 5  & 6 & 7 & 8 & 9 & $\geq$ 10 \\ \midrule
\multicolumn{7}{c}{NY}                                  \\ \midrule
BUR  & 4988  & 1217  & 401  & 120  & 48  & 16  & 8  & 69   \\ 
ROB  & 6153  & 1885  & 738  & 340  & 166  & 87  & 48  & 109    \\ 
TFT  & 20091  & 9628  & 5448  & 3420  & 2217  & 1538  & 1153  & 4412  \\ \midrule
\multicolumn{7}{c}{DC}                                  \\ \midrule
BUR  & 2178  & 600  & 206  & 94  & 43  & 17  & 8  & 6 \\ 
ROB  & 1359  & 436  & 168  & 77  & 36  & 18  & 12  & 11 \\ 
TFT  & 10564  & 5025  & 2871  & 1922  & 1349  & 1037  & 796  & 4180 \\ \midrule 
\multicolumn{7}{c}{CHI}                                 \\ \midrule
BUR  & 5851  & 1227  & 253  & 51  & 12  & 5  & 1 & 0\\ 
ROB  & 2886  & 811  & 290  & 101  & 40  & 16  & 5  & 4 \\ 
TFT  & 32345  & 14418  & 7815  & 4912  & 3299  & 2410  & 1885  & 9683 \\ \bottomrule
\end{tabular}
\end{table}

The computational time results is an excerpt on TABLE~\ref{table_time}, in the table, the load is the time for loading spatial-temporal information in CSV format. The R-tree is the time for building the R-tree index. The edges is the time to build edges based on R-tree, CC is the time to calculate connected components, the truss is the time to calculate k-truss, the core is the time to calculate k-core, the dbscan is the time to calculate k-DBSCAN. BGL is the time for the k-clique calculation using the boost graph library. No matter the size of the data, the dominant computational time is spent on the cohesive subgraph calculation.  It is observed that when the graph is small and less dense,  it takes less time to utilize BGL to compute graph properties.  In case the graph size is expanding fast, with larger clusters, it takes a considerable amount of time to get the k-clique result. Consequently, the advantage of approximate cohesive subgraph computation will be distinct. It seems our implementation is slower in the small graph case. Nevertheless, in general, the less cohesive the requirement of the results, the less time it takes to compute the result.

\subsubsection{Results comparison}

TABLES~\ref{table_core},~\ref{table_dbscan},~\ref{table_truss},~\ref{table_clique} show the distribution of the number of the cohesive subgraphs. Figure~\ref{coeshow} shows the clustering coefficient of the cohesive subgraphs detected using different methods on different data. We exclude the results of k-clique because the clustering coefficient is always 1. Although there are some variations of the results, we can, in general, conclude that k-truss is better than the k-DBSCAN, which is better than the k-core method. It is also worth noting that in many results when it comes to the cohesive subgraphs of large k (approximately $k>10$), subgraphs detected by the k-DBSCAN and k-core algorithm have very stable clustering coefficients, which might indicate some specific and stable graph patterns detected by these algorithms when $k$ is large.

\section{Conclusion}
In this paper, we have designed a Mapreduce based Knox test algorithm to help to prove the existence of a near-repeat effect on big data. We explore to identify efficient algorithms to derive near-repeat event chains. By representing crime events into a graph enabled by R-tree indexing, the near repeat crime chains can be derived through cohesive subgraph analysis. Four different cohesive subgraph analysis methods are implemented using AWS resources and compared concerning time and quality. The proposed solution has never been applied in the prior works on crime analysis and will have a broader impact on this research front in the future. However, there are still potential improvements to be made.

To begin with, we should perform the Knox test using event chain numbers such that we will be able to know whether the near-repeat effect also exists in the crime clusters. Meanwhile, we have noticed that there are the sheer amount of duplicated events in the real-world data, while the tightness of relationships between each event pair is not the same. How to handle these conditions in the sense of weighted vertex and edges is a challenging theoretical problem. Besides, when it comes to the larger amount of data, for example, to handle the online crime events, the data size will be much larger than what we process now. Hence, a parallel event chain detection algorithm is also needed. Since the crime events are adding each day, how to dynamically detecting event chains incrementally becomes an issue theoretically and practically. Last but not least, the near-repeat effect is not only existed in crime analysis but also existed in many areas such as transportation, how to utilize our method in other areas is a fascinating open problem.



%
%
%

\bibliographystyle{IEEEtran}
\bibliography{bare_conf}

\end{document}